\documentclass[aps,pra,twocolumn,superscriptaddress,showpacs]{revtex4}

\usepackage{graphics}%

\begin{document}


\title{Multichannel quantum-defect theory for 
	slow atomic collisions}


\author{Bo Gao}
\email[]{bgao@physics.utoledo.edu}
\homepage[]{http://bgaowww.physics.utoledo.edu}
\affiliation{Department of Physics and Astronomy,
	University of Toledo,
	Toledo, Ohio 43606}

\author{Eite Tiesinga}
\author{Carl J. Williams}
\author{Paul S. Julienne}
\affiliation{Atomic Physics Division,
	National Institute of Standards and Technology,
	Gaithersburg, Maryland 20899}


\date{\today}

\begin{abstract}

We present a multichannel quantum-defect theory for slow atomic collisions 
that takes advantages of the analytic solutions for the long-range potential,
and both the energy and the angular-momentum insensitivities
of the short-range parameters. The theory
provides an accurate and complete account of scattering processes, 
including shape and Feshbach resonances, in terms of a few parameters 
such as the singlet and the triplet scattering lengths.
As an example, results for $^{23}$Na-$^{23}$Na scattering are presented 
and compared close-coupling calculations.
 
\end{abstract}


\maketitle

\section{Introduction}
Slow atomic collisions are at the very foundation of cold-atom physics,
since they determine how atoms interact with each other and how this
interaction might be manipulated \cite{stw76,tie93}.
While substantial progress has been made over the past decade \cite{wei99},
there are still areas where the existing theoretical framework is
less than optimal. For example, all existing numerical methods 
may have difficulty with numerical stability in treating ultracold 
collisions in partial waves other than the $s$ wave, because the 
classically forbidden region grows infinitely wide as one approaches 
the threshold. This difficulty becomes a serious issue when there 
is a shape resonance right at or very close to the threshold, 
as the usual argument that the $s$ wave scattering dominates 
would no longer be applicable. Another area where a more optimal 
formulation is desirable is analytic representation.
Since much of our interest in cold atoms is in complex 
three-body and many-body physics, a simple, preferably analytical 
representation of cold collisions
would not only be very helpful to experimentalists, but also make it much 
easier to incorporate accurate two-body physics in
theories for three and many-atom systems.
Existing formulations of cold collisions provide little 
analytical results especially in cases, such as the alkali-metal atoms, where 
the atomic interaction is complicated by hyperfine structures.
Furthermore, whatever analytic results that we do have have been 
based almost exclusively on the effective-range
theory \cite{bla49}, the applicability of which is severely limited 
by the long-range atomic interaction \cite{gao98b,gao04c}.

Built upon existing multichannel quantum-defect theories 
that are based either on free-particle reference functions 
or on numerical solutions for the long-range 
potential \cite{mie80,jul89,gao96,bur98,mie00,rao04},
we present here a multichannel, angular-momentum-insensitive, quantum-defect 
theory (MAQDT) that overcomes many of the limitations of existing formulations.
It is a generalization of its single-channel counterpart 
\cite{gao01,gao00,gao04b}, and takes full advantage of both the analytic 
solutions for the long-range potential \cite{gao98a,gao99a}, 
and the angular momentum insensitivity of a properly defined short-range 
K matrix $K^c$ \cite{gao01,gao04b}. We show that as far as $K^c$ is 
concerned, the hyperfine interaction can be ignored, and the frame 
transformation \cite{rau71,lee73,lee75,gao96,bur98} applies basically exactly.
This conclusion \textit{greatly} simplifies the description
of any atomic collision that involves hyperfine structures. 
In the case of a collision between any two alkali-metal atoms in their 
ground state, whether they are identical 
or not, it reduces a complex multichannel problem to two 
\textit{single channel} problems. This property, along with the energy 
and angular-momentum insensitivity of $K^c$ \cite{gao01,gao04b}, 
leads to an accurate and \textit{complete} characterization
of slow collisions between any two alkali-metal atoms,
including shape resonances, Feshbach resonances, practically all
partial waves of interest, and over an energy-range of hundreds of
millikelvins, by four parameters for atoms with identical nuclei,
and five parameters for different atoms or different isotopes of 
the same atom. To be more specific, the four parameters can be taken as
the singlet $s$ wave scattering length $a_{0S}$, the triplet $s$ wave 
scattering length $a_{0T}$, the $C_6$ coefficient for the long-range
van der Waals potential $-C_6/r^6$, and the atomic hyperfine splitting 
$\Delta E^{HF}_a$ (The reduced mass $\mu$, which is also needed, is
not counted as a parameter since it is always fixed and well-known). 
For different atoms or different isotopes of 
the same atom, we need another hyperfine splitting for a total
of five parameters. These results also prepare us for future analytic 
representations of multichannel cold collisions, when we restrict 
ourselves to a smaller range of energies.

\section{MAQDT}
An $N$-channel, two-body problem can generally be described by a set of
wave functions
\begin{equation}
\psi_j = \sum_{i=1}^N \Phi_i F_{ij}(r)/r \;,
\label{eq:mchwfn}
\end{equation}
Here $\Phi_i$ are the channel functions describing 
all degrees of freedom other than the inter-particle distance 
$r$; and $F_{ij}(r)$ satisfies a set of close-coupling
equations
\begin{equation}
\left[-\frac{\hbar^2}{2\mu}\frac{d^2}{dr^2} 
	+\frac{\hbar^2 l_i(l_i+1)}{2\mu r^2}-E \right]F_{ij}
	+\sum_{j=1}^N V_{ij}(r)F_{ij} = 0 \;,
\label{eq:cceq}
\end{equation}
where $\mu$ is the reduced mass; $l_i$ is the relative angular momentum
in channel $i$; $E$ is the total energy; and $V_{ij}$ is the representation 
of inter-particle potential in the set of chosen channels
(see, e.g., reference \cite{gao96} for a diatomic system with hyperfine 
structures).

Consider now a class of problems for which the potential
at large distances ($r\ge r_0$) is of the form of
\begin{equation}
V_{ij}(r) = (E_i-C_{n_i}/r^{n_i})\delta_{ij} \;,
\label{eq:Vlarger}
\end{equation}
in the fragmentation channels that diagonalize the long-range interactions. 
Here $n_i>2$, and $E_i$ is the threshold energy associated with
a fragmentation channel $i$. 
As an example, for the scattering of two alkali-metal atoms in their
ground state, the fragmentation channels in the absence of any external
magnetic field are characterized by
the $FF$ coupling of reference \cite{gao96};
differences in threshold energies originate from atom hyperfine
interaction;
$n_i=6$ corresponds to the van der Waals interaction;
and $r_0$, with an order of magnitude around 30 a.u., corresponds
to the range of exchange interaction.

Before enforcing the physical boundary
condition (namely the condition that a wave function has to be 
finite everywhere) at infinity, Eqs.~(\ref{eq:cceq}) have $N$ 
linearly independent 
solutions that satisfy the boundary conditions at the origin.
For $r\ge r_0$, one set of these solutions can be written as
\begin{equation}
\psi^c_j = \sum_{i=1}^N \Phi_i
		(f^c_i\delta_{ij}-g^c_i K^{c}_{ij})/r \;.
\label{eq:Kcfch}
\end{equation}
Here $f^c_i$ and $g^c_i$ are the reference
functions for the long-range potential,
$-C_{n_i}/r^{n_i}$, in channel $i$, at energy 
$\epsilon_i=E-E_i$.
They are chosen such that they are 
independent of both the channel kinetic energy $\epsilon_i$ and 
the relative angular momentum $l_i$ at distances much smaller 
than the length scale $\beta_{n_i}=(2\mu C_{n_i}/\hbar^2)^{1/(n_i-2)}$ 
associated with the long-range interaction 
(see Appendix~\ref{sec:aqdtdefs} and references \cite{gao01,gao04b}).

Equation~(\ref{eq:Kcfch}) defines the short-range K matrix $K^c$.
It has a dimension equal to the total number of channels, $N$,
and encapsulates all the short-range physics. The $K^c$ matrix can either 
be obtained from numerical calculations (see Appendix~\ref{sec:Kcnum})
or be inferred from
other physical quantities such as the singlet and the triplet
scattering lengths, as discussed later in the article.

At energies where all $N$ channels are open, the solutions given by
Eq.~(\ref{eq:Kcfch}) already satisfy the physical boundary conditions
at infinity. Using the asymptotic behaviors of reference functions 
$f^c$ and $g^c$ at large $r$ (see Appendix~\ref{sec:aqdtdefs} 
and reference \cite{gao98a}), 
it is easy to show from Eq.~(\ref{eq:Kcfch}) that the physical 
$K$ matrix, defined by Eqs.~(4) and (5) of
reference \cite{gao96}, is an $N\times N$ matrix given 
in terms of $K^c$ by
\begin{equation}
K(E) = -( Z^c_{fg}-Z^c_{gg}K^{c} )
	(Z^c_{ff} - Z^c_{gf}K^{c})^{-1} \;.
\label{eq:Kphyo}	
\end{equation}
Here $Z^c_{fg}, Z^c_{gg}, Z^c_{ff}$, and $Z^c_{gf}$ are $N\times N$ 
diagonal matrices with diagonal elements given by
$Z^{c(n_i)}_{fg}(\epsilon_i, l_i), Z^{c(n_i)}_{gg}(\epsilon_i, l_i), 
Z^{c(n_i)}_{ff}(\epsilon_i, l_i)$, and $Z^{c(n_i)}_{gf}(\epsilon_i, l_i)$, 
respectively (see Appendix~\ref{sec:aqdtdefs} and 
references \cite{gao98a,gao00}). Equation~(\ref{eq:Kphyo}) is of the 
same form as its single channel counterpart 
\cite{gao01,gao00}, except that the relevant quantities are now
matrices, and $K^c$ is generally \textit{not} diagonal.

At energies where $N_o$ of the channels are open ($\epsilon_i>0$, for $i\in o$),
and $N_c=N-N_o$ of the channels are closed ($\epsilon_i<0$, for $i\in c$),
the physical boundary conditions at infinity leads to $N_c$ conditions that 
reduce that number of linearly independent
solutions to $N_o$ \cite{sea83,gao96,bur98}.
The asymptotic behavior of these $N_o$ solutions
gives the $N_o\times N_o$ physical $K$ matrix
\begin{equation}
K(E) = -( Z^c_{fg}-Z^c_{gg}K^{c}_{eff} )
	(Z^c_{ff} - Z^c_{gf}K^{c}_{eff})^{-1} \;.
\label{eq:Kphy}	
\end{equation}
Here $Z^c_{fg}, Z^c_{gg}, Z^c_{ff}$, and $Z^c_{gf}$ are $N_o\times N_o$ 
diagonal matrices with diagonal elements given by the corresponding
$Z^c$ matrix element for all open channels;
and we have defined the effective $K^c$ matrix for the open channels,
$K^c_{eff}$, to be
\begin{equation}
K^c_{eff} = K^{c}_{oo}+K^{c}_{oc}(\chi^c - K^{c}_{cc})^{-1}K^{c}_{co} \;.
\label{eq:Kceff}
\end{equation}
Here $\chi^c$ is an $N_c\times N_c$ diagonal 
matrix with elements $\chi^{c(n_i)}(\epsilon_i, l_i)$ 
(see Appendix~\ref{sec:aqdtdefs} and references \cite{gao98a,gao01})  
for all closed channels.
$K^{c}_{oo}$, $K^{c}_{oc}$, $K^{c}_{co}$, $K^{c}_{cc}$,
are submatrices of $K^{c}$ corresponding to open-open, open-closed,
closed-open, and closed-closed channels, respectively. 

All on-the-energy-shell scattering properties can be derived from 
the physical $K$ matrix. In particular, the physical 
$S$ matrix is given by \cite{gao96}
\begin{equation}
S(E) = [I+iK(E)][I-iK(E)]^{-1} \;,
\label{eq:smatrix}
\end{equation}
where $I$ represents a unit matrix.
From the $S$ matrix, the scattering amplitudes, the differential cross
sections, and other physical observables associated with scattering
can be easily deduced \cite{gao96}. 

It is worth noting that Eq.~(\ref{eq:Kphy}) preserve the form of
Eq.~(\ref{eq:Kphyo}). Thus the effect of closed channels is simply
to introduce an energy dependence, through $\chi^c$,
into the effective $K^c$ matrix, $K^c_{eff}$, for the open channels.
In particular, the bare (unshifted) locations of Feshbach resonances, 
if there are any, are determined by the solutions of
\begin{equation}
\det[\chi^c(E) - K^{c}_{cc}]=0 \;.
\label{eq:bound}
\end{equation}
They are locations of would-be bound states if the closed channels
are not coupled to the open channels. The same equation also gives
the bound spectrum of true bound states, at energies where all channels
are closed.

This completes our summary of MAQDT. It is completely rigorous
with no approximations involved. The theory is easily incorporated into
any numerical calculations (see Appendix~\ref{sec:Kcnum}). 
The difference from the standard approach
is that one matches the numerical wave function to the solutions of
the long-range potential to extract $K^c$, instead of matching to the
free-particle solutions to extract $K$ directly.
This procedure converges at a much smaller $r=r_0$,
the range of the exchange interaction, 
than methods that match to the free-particle solutions. 
Furthermore, since the propagation of the wave
function from $r_0$ to infinity is done analytically, through the $Z^c$ matrix
for open channels and $\chi^c$ function for closed channels, there is
no difficulty in treating shape resonances right at or very close to the
threshold. This improved convergence and stability
does not however fully illustrate the power of MAQDT formulation and
is not the focus of this article.
Instead, we focus here on the simple parameterization of slow atomic
collisions with hyperfine structures made possible by MAQDT.
The result also lays the ground work for future analytic representations 
of cold collisions.

\section{Simplified parameterization with frame transformation}

Equations~(\ref{eq:Kphyo})-(\ref{eq:Kceff}), and (\ref{eq:bound})
already provide a parameterization of slow-atom collisions and
diatomic bound spectra in terms of the elements of the $K^c$ matrix.
For alkali-metal atoms in their ground state, where the multichannel
nature arises from the hyperfine interaction, or a combination of
hyperfine and Zeeman interactions for scattering in a magnetic field, 
this parameterization
can be simplified much further by taking advantage of a frame 
transformation \cite{rau71,lee73,lee75,gao96,bur98}. 

At energies comparable to, or smaller than the atomic hyperfine
and/or Zeeman splitting, one faces the dichotomy that the hyperfine 
and/or Zeeman interaction,
while weak compared to the typical atomic interaction energy,
is sufficiently strong that the physical K matrix changes significantly
over a hyperfine splitting. (This is reflected in the very existence 
of Feshbach resonances \cite{stw76,tie93} and states with binding energies 
comparable to or small than the hyperfine splitting.)
As a result, the frame transformation does not apply directly to
the physical K matrix itself, and is generally a bad approximation even 
for the $K^0$ matrix of reference \cite{gao96}.
It was this recognition that first motivated the solutions
for the long-range potentials \cite{gao98a,gao99a}.

This dichotomy is easily and automatically resolved with the 
introduction of the short range K matrix $K^c$. The solution is simply to 
ignore the hyperfine and/or Zeeman interaction only at small distances 
and treat it exactly at large distances.
For $r<r_0$, the atomic interaction is of the order of the typical electronic
energy. Thus as far as $K^c$, which converges at $r_0$, is concerned,
the hyperfine and/or Zeeman interaction can be safely ignored. 
In this approximation, the $K^c$ matrix in the fragmentation channels 
can be obtained from the $K^c$ matrix in the condensation channels, 
namely the channels that diagonalize the short-range
interactions, by a frame transformation.

For simplicity, we restrict ourselves here to the case of 
zero external magnetic field, although the theory can readily be generalized to 
include a magnetic field.  The fragmentation channels
are the $FF$ coupled channels characterized by quantum numbers \cite{gao96}:
\[
(\alpha_1L_1S_1J_1I_1F_1)_A(\alpha_2L_2S_2J_2I_2F_2)_BFlTM_T \;,
\]
where $F$ results from the coupling of $F_1$ and $F_2$; $l$ is the relative
orbital angular momentum of the center-of-masses of the two atoms.
$T$ represents the total angular
momentum, and $M_T$ is its projection on a space-fixed axis \cite{gao96}.

Provided that the \textit{off-diagonal} second-order spin-orbital
coupling \cite{mie96} can be ignored, a good approximation for
lighter alkali-metal atoms, or more generally, for any physical
processes that are allowed by the exchange interaction,
the condensation channels can be taken as 
the $LS$ coupled channels characterized by quantum numbers \cite{gao96}:
\[ 
	(\alpha_1L_1S_1I_1)_A(\alpha_2L_2S_2I_2)_BLl{\cal L}SKITM_T \;, 
\] 
where $\mathbf{\cal L}=\mathbf{L}+\mathbf{l}$ is the total orbital 
angular momentum. $\mathbf{S}=\mathbf{S}_1+\mathbf{S}_2$ is the
total electron spin. $\mathbf{I}=\mathbf{I}_1+\mathbf{I}_2$ is the total
nuclear spin. And $\mathbf{K}=\mathbf{\cal L}+\mathbf{S}$ is the total 
angular momentum excluding nuclear spin. 

Ignoring hyperfine interactions, as argued earlier,
the $K^c$ matrix in $FF$-coupled channels,
labeled by index $i$ or $j$, is related to the $K^c$ matrix
in $LS$-coupled channels, labeled by index $\alpha$ or $\beta$, 
by a frame transformation \cite{gao96}
\begin{equation}
K^{c}_{ij} = \sum_{\alpha\beta}
	U_{i\alpha}K^{c(LS)}_{\alpha\beta}U_{j\beta} \;,
\label{eq:ftK}
\end{equation}
where $K^{c(LS)}$ is the $K^c$ matrix computed in the LS coupling
\textit{with the hyperfine interactions ignored}.
The most general form of frame transformation 
$U_{j\beta}$ is given by Eq.~(49) of reference \cite{gao96}.
For collision between any two atoms with zero orbital angular momentum,
$L_1=L_2=L=0$, including of course any two alkali-metal atoms in their ground states,
the frame transformation simplifies to
\begin{eqnarray}
U_{i\beta}(T) &=& \delta_{l_il_\beta}
	(-1)^{F_i+S_\beta+I_\beta}
	[F_{1i}, F_{2i}, F_i, 
	S_\beta, K_\beta, I_\beta]^{1/2} \nonumber\\
	& & \times
	\left\{\begin{array}{ccc} F_i & l_i & T \\ 
	K_\beta & I_\beta & S_\beta \end{array} \right\} 
    \left\{\begin{array}{ccc} S_{1} & S_{2} & S_\beta \\
	I_1 & I_2 & I_\beta \\ F_{1i} & F_{2i} & F_i \end{array}\right\} \;,
\label{eq:lstoff}
\end{eqnarray}
for atoms with different nuclei. 
Here $[a,b,\dots]\equiv(2a+1)(2b+1)\cdots$.
For two atoms with identical nuclei,
the same transformation needs to be multiplied by a normalization
factor \cite{gao96}
\begin{eqnarray}
U_{\{i\}\{\beta\}} &=& \{ 1+\delta(\alpha_2L_2S_2,\alpha_1L_1S_1)
	[1-\delta(J_{2i}F_{2i},J_{1i}F_{1i})] \}^{1/2}\nonumber\\
	& &\times U_{i\beta}\;.
\label{eq:lstoffin}
\end{eqnarray}

We emphasize that to the degree that the hyperfine interaction in a slow atomic
collision can be approximated by \textit{atomic} hyperfine interactions,
as has always been assumed \cite{tie93}, the frame transformation given 
by Eq.~(\ref{eq:ftK}) should be regarded as \textit{exact}. If the hyperfine 
interaction inside $r_0$, the range of the exchange interaction, cannot
be ignored, the true molecular hyperfine interaction \cite{bab91} 
would have to be used. Inclusion of \textit{atomic} hyperfine 
interactions inside $r_0$ is simply another approximation, and an 
unnecessary complication, that is 
of the same order of accuracy as ignoring it 
completely. In other words, any real improvement over the frame 
transformation has to require a better treatment 
of \textit{molecular hyperfine interactions} \cite{bab91}.
A similar statement is also applicable to the Zeeman interaction.

The applicability of the frame transformation greatly simplifies
the description of any slow atomic collision with hyperfine structures.
For alkali-metal atoms in their ground state, 
and ignoring off-diagonal second-order spin-orbital
coupling \cite{mie96},
it reduces a complex multichannel problem  to two single
channel problems, one for the singlet $S=0$, and one for the
triplet, $S=1$, with their respective single-channel $K^c$ \cite{gao01,gao00} 
denoted by $K^c_S(\epsilon,l_i)$ and $K^c_T(\epsilon,l_i)$, respectively.
The $K^c$ matrix in the LS coupling, $K^{c(LS)}$, is diagonal with 
diagonal elements given by either $K^c_S$ or $K^c_T$ \cite{gao96}.
Ignoring the energy and the angular momentum dependences
of $K^c_S(\epsilon,l_i)$ and $K^c_T(\epsilon,l_i)$ \cite{gao01,gao04b},
they become simply two parameters $K^c_S=K^c_S(0,0)$ and $K^c_T=K^c_T(0,0)$,
which are related to the singlet and the triplet $s$ wave
scattering lengths by \cite{gao03a}
\begin{equation}
a_{0}/\beta_n = \left[b^{2b}\frac{\Gamma(1-b)}{\Gamma(1+b)}\right]
	\frac{K^c(0,0) + \tan(\pi b/2)}{K^c(0,0) - \tan(\pi b/2)} \;,
\label{eq:a0sKc}
\end{equation}
where $b=1/(n-2)$ with $n=6$ for alkali-metal scattering in the ground
state. With $K^{c(LS)}$, and therefore $K^c$, being parameterized 
by two parameters, a complete
parameterization of alkali-metal scattering requires only two, or three, 
more parameters including $C_6$, which determined the length and
energy scales for the long range interaction, and the atomic
hyperfine splitting $\Delta E^{HF}_a$, which characterizes the 
strength of atomic hyperfine interaction and also determines the 
channel energies.

We note here that our formulation ignores
the weak magnetic dipole-dipole interaction \cite{sto88,mie96}. 
It is important only for processes, such as the dipolar
relaxation, that are not allowed by the exchange interaction.
Such processes can be incorporated perturbatively 
after a MAQDT treatment \cite{mie00}. 
We also note that for processes, such as the
spin relaxation of Cs, for which the off-diagonal second-order 
spin-orbital coupling is important \cite{mie96,leo00}, 
a different choice of condensation channels, similar to
the $JJ$-coupled channels of reference \cite{gao96}, 
would be required. 
The resulting description is similar conceptually, but 
involves more parameters \cite{mie96,leo00}. 

\section{Sample results for sodium-sodium scattering}

\begin{figure}
\scalebox{0.38}{\includegraphics{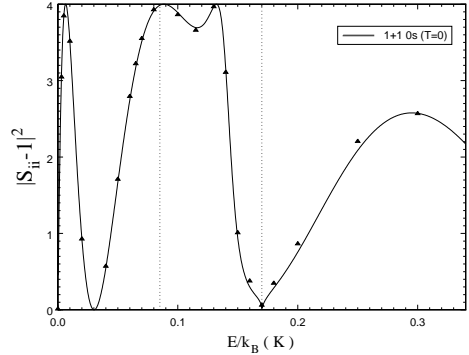}}
\caption{$|S_{ii}-1|^2$, where $S_{ii}$ is an $S$ matrix element, 
for the $s$ wave elastic scattering of two $^{23}$Na atoms in channel 
[$\{F_1=1,F_2=1\}F=0,l=0,T=0$], as a function of $E/k_B$, where
$k_B$ is the Boltzmann constant. The vertical lines identify the 
locations of thresholds for $\{F_1=1,F_2=2\}$ and $\{F_1=2,F_2=2\}$ 
channels. Solid line: results of a four-parameter MAQDT parameterization.
Points: results of close-coupling calculations.
\label{Figure1}}
\end{figure}

\begin{figure}
\scalebox{0.38}{\includegraphics{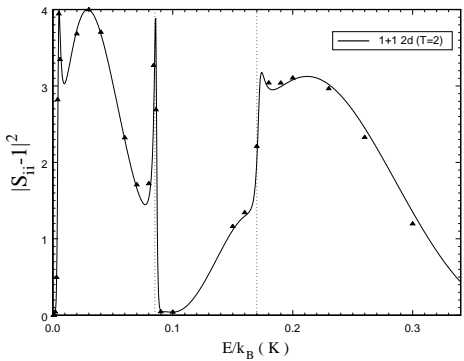}}
\caption{The same as Figure~1 excepts it is for $d$ wave channel
[$\{F_1=1,F_2=1\}F=2,l=2,T=2$].
\label{Figure2}}
\end{figure}

\begin{figure}
\scalebox{0.38}{\includegraphics{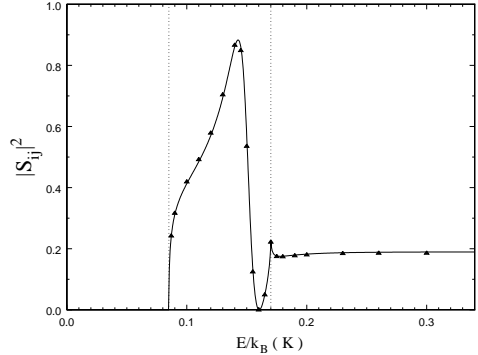}}
\caption{The $S$ matrix element, $|S_{ij}|^2$, for the $s$ wave 
inelastic scattering of two $^{23}$Na atoms between channels 
[$\{F_1=1,F_2=1\}F=2,l=0,T=2$] and [$\{F_1=1,F_2=2\}F=2,l=0,T=2$].
\label{Figure3}}
\end{figure}

As an example, Figures~\ref{Figure1}-\ref{Figure3} show the
comparison between close-coupling calculations and a four-parameter 
MAQDT parameterization for slow atomic collision between a pair
of $^{23}$Na atoms in the absence of exteranl magnetic field. 
The points are the close-coupling results 
using the potentials of references \cite{sam00,lau02}.
The curves represent the results of a four-parameter 
parameterization with $a_{0S}=19.69$ a.u., $a_{0T}=64.57$ a.u.,
$C_6 = 1556$ a.u. \cite{der99}, and 
$\delta E^{HF}_a = 1772$ MHz,
where $a_{0S}$ and $a_{0T}$ are computed from the singlet and
the triplet potentials of references \cite{sam00,lau02}.
Figure~1 shows the $S$ matrix element for the $s$ wave elastic 
scattering in channel [$\{F_1=1,F_2=1\}F=0,l=0,T=0$]. 
The feature around 130 mK is a Feshbach resonance in 
channel [$\{F_1=2,F_2=2\}F=0,l=0,T=0$].
For this particular case, $K^{c(LS)}$ is a $2\times 2$ matrix
\begin{equation}
K^{c(LS)} = \left( \begin{array}{cc}
	K^{c}_S & 0 \\
	0 & K^{c}_T \end{array} \right) \;,
\end{equation}
with channel ordering shown in Table~\ref{tb:schs}.
\begin{table}
\caption{Channel structure for $s$ wave scattering between
	two identical atoms with $L_1=L_2=0$, $S_1=S_2=1/2$, 
	and $I_1=I_2=3/2$, in the absence of external magnetic field. 
	Examples include $^7$Li, $^{23}$Na, $^{39}$K and $^{87}$Rb. 
	\label{tb:schs}}
\begin{ruledtabular}
\begin{tabular}{c|c|c}
$T$ & LS coupling $(S,I)$ & FF coupling $\{F_1,F_2\}F$ \\
\hline
0 & S=0, I=0 & \{1,1\}0 \\
  & S=1, I=1 & \{2,2\}0 \\
\hline
1 & S=1, I=1 & \{1,2\}1 \\
\hline
2 & S=0, I=2 & \{1,1\}2 \\
  & S=1, I=1 & \{1,2\}2 \\
  & S=1, I=3 & \{2,2\}2 \\
\hline
3 & S=1, I=3 & \{1,2\}3 \\
\hline
4 & S=1, I=3 & \{2,2\}4 \\
\end{tabular}
\end{ruledtabular}
\end{table}
$K^{c}_S$ and $K^{c}_T$ are related to the singlet and the
triplet scattering lengths by Eq.~(\ref{eq:a0sKc}).
The frame transformation is given by 
[c.f. Eqs.~(\ref{eq:lstoff}) and (\ref{eq:lstoffin})]
\begin{equation}
U(T=0) = \frac{1}{2\sqrt{2}}
	\left( \begin{array}{cc}
	\sqrt{3} & \sqrt{5} \\
	\sqrt{5} & -\sqrt{3} 
	\end{array} \right) \;,
\end{equation}
which leads to
\begin{equation}
K^{c} = \frac{1}{8}\left( \begin{array}{cc}
	3K^{c}_S+5K^{c}_T & \sqrt{15}\left(K^{c}_S-K^{c}_T\right) \\
	\sqrt{15}\left(K^{c}_S-K^{c}_T\right) & 5K^{c}_S+3K^{c}_T 
	\end{array} \right) \;.
\label{eq:Kc110s0}	
\end{equation}
From the $K^c$ matrix, the $S$ matrix is obtained from the MAQDT
equations~(\ref{eq:Kphyo})-(\ref{eq:smatrix}). Note how 
Eq.~(\ref{eq:Kc110s0}) shows explicitly that the off-diagonal
element of $K^c$, which determines the rate of inelastic
collision due to exchange interaction, goes to zero
for $K^{c}_S=K^{c}_T$, namely when $a_{0S}=a_{0T}$.

The results presented in Figs.~2 and 3 are obtained in similar fashion.
Figure~2 shows the $S$ matrix element for the $d$ wave elastic 
scattering in channel [$\{F_1=1,F_2=1\}F=2,l=2,T=2$].
It illustrates how the same parameters that we use to
describe the $s$ wave scattering also describe the $d$ wave
scattering, due to the fact that $K^c_S$ and $K^c_T$ 
are insensitive to $l$ \cite{gao01,gao04b}.
Here the sharp features around the thresholds
are $d$ wave shape resonances.
Figure~3 shows the $S$ matrix element for the $s$ wave inelastic 
scattering between channel [$\{F_1=1,F_2=1\}F=2,l=0,T=2$] and
channel [$\{F_1=1,F_2=2\}F=2,l=0,T=2$].
The kinks (discontinuities in the derivative), in both Fig.~\ref{Figure3} and 
Fig.~\ref{Figure1} at the $\{F_1=2,F_2=2\}$ threshold,
are general features associated with the opening of an $s$ wave channel.
There is no kink at the $\{F_1=1,F_2=2\}$ threshold in Figure~\ref{Figure1}
because the [$\{F_1=1,F_2=1\}F=0,l=0,T=0$] channel is not coupled to
$\{F_1=1,F_2=2\}$ channels.

The agreements between the MAQDT parameterization and close-coupling
calculations are excellent, exact for all practical purposes,
in all cases. Conceptually, these results illustrate that
through a proper MAQDT formulation,
atomic collision over a wide range of energies (300 mK compared
to the Doppler cooling limit of about 0.2 mK for $^{23}$Na), with complex
structures including Feshbach and shape resonances, and for different
partial waves, can all be described by parameters that we often associate
with the $s$ wave scattering at zero energy only, namely the singlet
and the triplet scattering lengths.

\section{Conclusion}

In conclusion, a multichannel, angular-momentum-insensitive, quantum-defect 
theory (MAQDT) for slow atomic collisions has been presented. 
We believe it to be the optimal formulation for purposes including
exact numerical calculation, parameterization, and analytic representation.
We have shown that by dealing with the short-range K matrix $K^c$,
the frame transformation becomes basically exact, which greatly simplifies 
the description of any slow atomic collision with hyperfine structures.
As an example, we have shown that even a simplest parameterization
with four parameters, in which the energy and the $l$ dependence
of $K^c_S$ and $K^c_T$ are completely ignored, reproduces the close-coupling
calculations for $^{23}$Na atoms over a wide range of energies
basically exactly. The effect of an external magnetic field, which is
not considered in this article, is easily incorporated as it simply requires 
another frame transformation \cite{bur98}. 

The concepts and the main constructs of the theory can be generalized to
other scattering processes including ion-atom scattering and
atom-atom scattering in excited states. The key difference will be in the
long-range interaction [c.f. Eq.~(\ref{eq:Vlarger})]. 
In addition to possibly different long-range exponent
$n_i$ (such as $n_i=4$ for ion-atom scattering), there may also be 
long-range off-diagonal
coupling that will have to be treated differently.

Finally, we expect that 
if we restrict ourselves to a smaller range of energies, 
of the order of $(\hbar^2/2\mu)(1/\beta_6)^2$ (about 1 mK for $^{23}$Na),
a number of analytic results, similar to the single-channel results of references 
\cite{gao98b} and \cite{gao04c}, can be derived even for the complex multichannel
problem of alkali-metal collisions.
These results may, in particular, lead to a more general and more 
rigorous parameterization of magnetic Feshbach resonances 
(see, e.g., references \cite{mar04,rao04} for some recent works in this area).

\begin{acknowledgments}
Bo Gao was supported by the National Science Foundation under 
the Grant number PHY-0140295.
\end{acknowledgments}

\appendix
\section{Definitions of MAQDT functions}
\label{sec:aqdtdefs}

The reference functions $f^c$ and $g^c$ for a $-C_n/r^n$ ($n>2$)
potential are a pair of 
linearly independent solutions of the
radial Schr\"{o}dinger equation 
\begin{equation}
\left[-\frac{\hbar^2}{2\mu}\frac{d^2}{dr^2} 
	+\frac{\hbar^2 l(l+1)}{2\mu r^2}-\frac{C_n}{r^n}-\epsilon \right]
	u_{\epsilon l}(r) = 0 \;,
\label{eq:invrnsch}
\end{equation}
which can be written in a dimensionless form as
\begin{equation}
\left[\frac{d^2}{dr_s^2} - \frac{l(l+1)}{r_s^2}
	+ \frac{1}{r_s^n} + \epsilon_s\right]
	u_{\epsilon_s l}(r_s) = 0 \;,
\label{eq:invrn}
\end{equation}
where $r_s = r/\beta_n$ is a scaled radius, 
$\beta_n \equiv (2\mu C_n/\hbar^2)^{1/(n-2)}$ is the length scale
associated with the $-C_n/r^n$ interaction, and
\begin{equation}
\epsilon_s = \frac{\epsilon}{(\hbar^2/2\mu)(1/\beta_n)^2} \;,
\label{eq:es}
\end{equation}
is a scaled energy.

The $f^c$ and $g^c$ pair are chosen such that they
have not only energy-independent, but also angular-momentum-independent 
behaviors in the region of $r\ll\beta_n$ (namely $r_s\ll 1$):
\begin{eqnarray}
f^c_{\epsilon_s l}(r_s) &\stackrel{r_s\ll 1}{\longrightarrow}& 
	(2/\pi)^{1/2}r_s^{n/4}
	\cos\left(y-\pi/4 \right) \;, 
\label{eq:fcasy0}\\
g^c_{\epsilon_s l}(r_s) &\stackrel{r_s\ll 1}{\longrightarrow}& 
	-(2/\pi)^{1/2}r_s^{n/4}
	\sin\left(y -\pi/4 \right) \;,
\label{eq:gcasy0}
\end{eqnarray}
for all energies \cite{gao01,gao03a}. Here $y=[2/(n-2)]r_s^{-(n-2)/2}$.
They are normalized such that 
\begin{equation}
W(f^c,g^c) \equiv f^c\frac{d g^c}{dr_s}-\frac{d f^c}{dr_s}g^c
	= 2/\pi \;.
\end{equation}

For $\epsilon=0$, the $f^c$ and $g^c$ pair for arbitrary $n$ can be
found in reference \cite{gao04b}.
For $\epsilon\neq 0$, the $f^c$ and $g^c$ pair for $n=6$ can be
found in reference \cite{gao04a}. The are related to the $f^0$ and
$g^0$ pair of reference \cite{gao98a} by
\begin{widetext}
\begin{equation}
\left( \begin{array}{c}
	f^c \\
	g^c \end{array} \right)
	= \frac{1}{\sqrt{2}}
	\left( \begin{array}{cc}
	\cos(\pi\nu_0/2) & \sin(\pi\nu_0/2) \\
	-\sin(\pi\nu_0/2) & \cos(\pi\nu_0/2) \end{array} \right)
	\left( \begin{array}{cc}
	1 & 0 \\
	0 & -1 \end{array} \right)
	\left( \begin{array}{c}
	f^0 \\
	g^0 \end{array} \right) \;,
\label{eq:c0trans}	
\end{equation}
where $\nu_0 = (2l+1)/4$ for $n=6$.

The $Z^{c(n)}(\epsilon_s,l)$ matrix is defined by the large-$r$ asymptotic 
behaviors of $f^c$ and $g^c$ for $\epsilon>0$
\begin{eqnarray}
f^c_{\epsilon_s l}(r_s) &\stackrel{r\rightarrow \infty}{\longrightarrow}&
	\sqrt{\frac{2}{\pi k_s}}\left[Z^{c(n)}_{ff}(\epsilon_s,l)
	\sin\left(k_sr_s-\frac{l\pi}{2}\right) 
	- Z^{c(n)}_{fg}(\epsilon_s,l)\cos\left(k_sr_s-\frac{l\pi}{2}\right)\right] \;,
\label{eq:fclarger}\\	 
g^c_{\epsilon_s l}(r_s) &\stackrel{r\rightarrow \infty}{\longrightarrow}&
	\sqrt{\frac{2}{\pi k_s}}\left[Z^{c(n)}_{gf}(\epsilon_s,l)
	\sin\left(k_sr_s-\frac{l\pi}{2}\right) 
	- Z^{c(n)}_{gg}(\epsilon_s,l)\cos\left(k_sr_s-\frac{l\pi}{2}\right)\right] \;,
\label{eq:gclarger}
\end{eqnarray}
\end{widetext}
where $k_s=k\beta_n$ with $k=(2\mu\epsilon/\hbar^2)^{1/2}$. 
This define a $2\times 2$ $Z^{c(n)}(\epsilon_s,l)$ matrix
\begin{equation}
Z^{c(n)} = \left( \begin{array}{cc}
	Z^{c(n)}_{ff} & Z^{c(n)}_{fg} \\
	Z^{c(n)}_{gf} & Z^{c(n)}_{gg} \end{array} \right) \;.
\end{equation}
It is normalized such that 
\begin{equation}
\det\left[Z^{c(n)}\right] = Z^{c(n)}_{ff}Z^{c(n)}_{gg}
	- Z^{c(n)}_{gf}Z^{c(n)}_{fg} = 1 \;.
\end{equation}

The $\chi^{c(n)}_l(\epsilon_s)$ function is defined through the large-$r$ 
asymptotic behaviors of $f^c$ and $g^c$ for $\epsilon<0$.
\begin{widetext}
\begin{eqnarray}
f^c_{\epsilon_s l}(r_s) &\stackrel{r\rightarrow \infty}{\longrightarrow}&
	(2\pi \kappa_s)^{-1/2}\left[W^{c(n)}_{f-}(\epsilon_s,l)e^{\kappa_s r_s} 
	- W^{c(n)}_{f+}(\epsilon_s,l)(2e^{-\kappa_i r})\right] \;, \\	 
g^c_{\epsilon_s l}(r_s) &\stackrel{r\rightarrow \infty}{\longrightarrow}&
	(2\pi \kappa_s)^{-1/2}\left[W^{c(n)}_{g-}(\epsilon_s,l)e^{\kappa_s r_s} 
	- W^{c(n)}_{g+}(\epsilon_s,l)(2e^{-\kappa_s r_s})\right] \;,
\end{eqnarray}
\end{widetext}
where $\kappa_s=\kappa\beta_n$ with $\kappa=(2\mu|\epsilon_i|/\hbar^2)^{1/2}$.
This defines a $2\times 2$ $W^{c(n)}(\epsilon_s,l)$ matrix, 
\begin{equation}
W^{c(n)} = \left( \begin{array}{cc}
	W^{c(n)}_{f-} & W^{c(n)}_{f+} \\
	W^{c(n)}_{g-} & W^{c(n)}_{g+} \end{array} \right) \;,
\end{equation}
from which the 
$\chi^{c(n)}_l(\epsilon_s)$ function is defined by
\begin{equation}
\chi^{c(n)}_l(\epsilon_s) = W^{c(n)}_{f-}/W^{c(n)}_{g-} \;.
\label{eq:chic}
\end{equation}
The $W^{c(n)}$ matrix is normalized such that 
\begin{equation}
\det\left[W^{c(n)}\right] = 
	W^{c(n)}_{f-}W^{c(n)}_{g+}-W^{c(n)}_{g-}W^{c(n)}_{f+} = 1 \;.
\end{equation}

The $Z^{c(n)}(\epsilon_s,l)$ and $W^{c(n)}(\epsilon_s,l)$ matrices, 
for $\epsilon>0$ and $\epsilon<0$, 
respectively, describe the propagation of
a wave function in a $-C_n/r^n$ potential from small to large distances,
or vice versa. They are universal functions of the scaled energy $\epsilon_s$
with their functional forms determined only by the exponent $n$ of the 
long-range potential and the $l$ quantum number. 
The $C_n$ coefficient and the reduced mass play a role
only in determining the length and energy scales.

The $Z^{c(n)}$ matrix for $n=6$ is given in reference \cite{gao00}.
The $\chi^{c(n)}_l(\epsilon_s)$ function for $n=6$ is given in
reference \cite{gao01}. They are derived from Eq.~(\ref{eq:c0trans}) and
the asymptotic behaviors of the $f^0$ and $g^0$ pair given in reference 
\cite{gao98a}. 

\section{$K^c$ from numerical solutions}
\label{sec:Kcnum}

Let $F(r)$ be the matrix, with elements $F_{ij}(r)$, representing any $N$ linearly
independent solutions of the close-coupling equation,
and $F^\prime(r)$ be its corresponding derivative
[Each column of $F(r)$ corresponds to one solution through Eq.~(\ref{eq:mchwfn})]. 
For $r\ge r_0$, $F$ can always be written as
\begin{equation}
F(r) = f^c(r)A-g^c(r)B \;,
\label{eq:Flarger}
\end{equation}
where $f^c(r)$ and $g^c(r)$ are $N\times N$ diagonal matrices with diagonal elements
given by $f^c_i(r)$ and $g^c_i(r)$, respectively.
The matrices $A$ and $B$ can be obtain, e.g., from knowing $F(r)$ and $F^\prime(r)$
at one particular $r\ge r_0$. Specifically
\begin{eqnarray}
A &=& (\pi\beta_n/2)[g^{c\prime}(r) F(r)-g^c(r) F^{\prime}(r)]\;, \\
B &=& (\pi\beta_n/2)[f^{c\prime}(r) F(r)-f^c(r) F^{\prime}(r)] \;.
\end{eqnarray}
Comparing Eq.~(\ref{eq:Flarger}) with Eq.~(\ref{eq:Kcfch}) gives
\begin{equation}
K^c = [f^{c\prime}(r) F(r)-f^c(r) F^{\prime}(r)]
	[g^{c\prime}(r) F(r)-g^c(r) F^{\prime}(r)]^{-1} \;.
\label{eq:Kcnum}	
\end{equation}
In an actual numerical calculation, which can be implemented using a number of
different methods \cite{raw99}, the right-hand-side (RHS) of this equation is evaluated at
progressively greater $r$ until $K^c$ converges to a constant matrix to
a desired accuracy. This procedure also provides a numerical definition
of $r_0$, namely it is the radius at which the RHS of Eq.(\ref{eq:Kcnum}) becomes
a $r$-independent constant matrix. 

\bibliography{sac}

\end{document}